\documentclass[twocolumn,showpacs,preprintnumber,amsmath,amssymb,aps,prb]{revtex4}
\usepackage{graphicx}
\begin{document}

\title{  
Commensurability Effects at Nonmatching Fields for Vortices in
Diluted Periodic Pinning Arrays}
\author{C.~Reichhardt and C.~J.~Olson Reichhardt} 
\affiliation{ 
Theoretical Division, 
Los Alamos National Laboratory, Los Alamos, New Mexico 87545}

\date{\today}
\begin{abstract}
Using numerical simulations,
we demonstrate that superconductors containing
periodic pinning arrays which have been diluted
through random removal of a fraction of the pins have
pronounced commensurability effects at the same magnetic field strength
as undiluted pinning arrays.  The commensuration can occur at fields
significantly higher than the matching field, produces much greater
critical current enhancement than a random pinning arrangement due to
suppression of vortex channeling,
and persists for 
arrays with up to 90\% dilution.  
These results 
suggest that diluted periodic pinning arrays may be a promising geometry
to increase the critical current in superconductors over a wide magnetic
field range.
\end{abstract}
\pacs{74.25.Qt}
\maketitle

\vskip2pc

\section{Introduction}

One of the most important issues 
for applications of 
type-II superconductors is 
achieving the highest possible critical current.
This requires preventing the depinning and motion of the 
superconducting
vortices that are present in the sample.
A promising approach is the use of artificial pinning arrays,
which have been the focus of extensive recent studies.
Pinning arrays with triangular, square, and rectangular geometries 
have been fabricated in superconductors 
using either microholes or blind holes 
\cite{Fiory,Baert,Harada,Metlushko,Bending,Karapetrov} 
or arrays of magnetic dots \cite{Schuller,Fertig}.
The 
resulting critical currents are significantly enhanced 
at the matching magnetic field $B_\phi$ where the number of 
vortices equals the number of pinning sites, as well as at higher
fields $nB_\phi$, with $n$ an integer.
At the matching fields,
peaks or anomalies in the critical current occur and 
the vortices form highly ordered commensurate lattices which are
free of topological defects 
\cite{Harada,Bending,Karapetrov,Fertig,Reichhardt,Jensen,Zimanyi,Olson,Peeters}.   
Although it 
might appear that the best method for increasing the overall critical current
would be to increase the pinning density and thus increase $B_\phi$,
this technique fails since adding too many pins to the sample
can degrade the overall
superconducting properties by lowering $T_{c}$ and 
$H_{c2}$.   
Thus, an open question is how the critical current 
can be maximized using the {\it smallest} possible number of
pinning sites. Since 
completely periodic pinning arrays have been shown to enhance pinning,  
one can ask whether other
pinning arrangements such as quasiordered or semiordered arrays could
be even more effective at pinning vortices.  
In recent simulations \cite{Misko} and experiments \cite{Koelle,Villegas},
it was demonstrated that a quasicrystalline pinning array produced an
enhancement of the critical current compared to a random pinning array for
fields below and up to $B_\phi$. 
The enhancement disappears for fields above $B_\phi$, however. 

In this work we 
demonstrate a new type of commensurability effect that occurs in 
periodic pinning arrays 
that have been
diluted by randomly removing 
a fraction of the pinning sites. 
In an undiluted array, the commensurability effects 
occur at fields $nB_\phi$.
In a diluted array,
$B_{\phi}$ is reduced; however, the pinning array still retains correlations
which are associated with the periodicity of the original 
undiluted pinning array.
In this case, in addition to commensuration effects at $B_\phi$, we
observe  
noticeable commensuration effects 
at the higher field $B^{*}_{\phi}$, the matching field
for the {\it undiluted} pinning array, even though many of the pins
that would have been present in such an array are missing.
Strong peaks in the critical current 
associated with well-ordered vortex configurations appear at $nB^{*}_\phi$.
This commensurability effect is remarkably robust 
and still appears in arrays which have up to 90\% 
of the pinning sites removed, so that $B^{*}_\phi=10B_\phi$. 
In samples with equal numbers of pinning sites, diluted 
periodic pinning arrays produce
considerable enhancement of the critical current
for fields 
well above $B_\phi$ compared to random pinning arrays.
The diluted periodic 
pinning arrays also have a reduced amount of one-dimensional vortex  
channeling above $B_\phi$ compared to undiluted periodic pinning arrays. 
Such easy vortex channeling 
is what leads to the reduction of the critical current 
above $B_\phi$ in the undiluted arrays. 
Our results should also 
apply to related systems such as colloids
interacting with periodic substrates \cite{Bechinger} and vortices in Bose-Einstein condensates interacting with
optical traps \cite{Bigelow}.

\section{Simulation}

We simulate a two-dimensional system of size $L_x \times L_y$
with periodic boundary conditions in the $x$ and $y$ directions
containing $N_{v}$ vortices and $N_{p}$ pinning sites. 
The vortices are modeled as point particles where the dynamics of 
vortex $i$ is governed by the following equation of motion:
\begin{equation}
\eta\frac{d {\bf R}_i}{dt} =  {\bf F}^{vv}_{i} + {\bf F}^{vp}_{i} +  {\bf F}^{L} .
\end{equation} 
The damping constant $\eta = \phi^{2}_{0}d/2\pi \xi^2 \rho_{N}$,
where $d$ is the sample thickness, $\xi$ is the coherence length,
$\rho_{N}$ is the normal-state resistivity, and $\phi_{0} = h/2e$ is the
elementary flux quantum. 
The vortex-vortex interaction force
is 
\begin{equation} 
{\bf F}^{vv}_{i} = \sum^{N_{v}}_{j \neq i}A_vf_{0}K_{1}(R_{ij}/\lambda){\bf {\hat R}}_{ij}
\end{equation} 
where $K_{1}$ is the modified Bessel function,
$A_v$ is an interaction prefactor which is set to 1 unless otherwise
noted, 
$\lambda$ is the London penetration
depth, $f_{0} = \phi^{2}_{0}/2\pi\mu_0\lambda^{3}$, 
${\bf R}_{i(j)}$ is the position of vortex $i(j)$,
$R_{ij}=|{\bf R}_{i}-{\bf R}_{j}|$, and 
${\bf {\hat R}}_{ij}=({\bf R}_i-{\bf R}_j)/R_{ij}$.
Forces are measured in units of $f_{0}$ and distances in units of $\lambda$.   
The vortex density $B=N_v/L_xL_y$. 
The vortex-vortex interaction force falls off sufficiently rapidly 
that a cutoff at $R_{ij}=6\lambda$ is imposed for
computational efficiency and a short range cutoff at 
$R_{ij}=0.1\lambda$ is also imposed to avoid
a force divergence. 
The 
$N_p$ pinning sites are modeled as attractive parabolic 
potential traps of radius $r_{p}=0.2\lambda$ and strength 
$f_{p}=0.25f_0$, with
${\bf F}^{vp}_{i} = (f_{p}/r_{p})R_{ik}\Theta((r_{p} - R_{ik})/\lambda){\bf {\hat r}}^{(p)}_{ik}$.
Here ${\bf R}_{k}^{(p)}$ is the location of pinning site $k$,
$R_{ik}=|{\bf R}_i-{\bf R}_k^{(p)}|$, 
${\bf {\hat R}}_{ik}=({\bf R}_i-{\bf R}_k^{(p)})/R_{ik}$, and 
$\Theta$ is the Heaviside step function. 
The pinning density is 
$B_\phi=N_p/(L_xL_y)$.
The pins are placed in a triangular 
array with matching field $B_\phi^*$ that 
has had a fraction $P_d$ of the pins removed,
so that $B_\phi=(1-P_d)B^{*}_\phi$.
The Lorentz driving force ${\bf F}^{L}=F^L{\bf {\hat x}}$, 
assumed uniform on all vortices, 
is generated by and perpendicular to an externally applied current 
${\bf J}$.  
The initial vortex positions are obtained by simulated annealing from
a high temperature.
After annealing, the temperature is set to zero and the
external drive $F^L$ is applied in 
increments of $2\times 10^{-5}f_{0}$ 
every $5\times 10^3$ simulation time steps. 
We measure the average 
vortex velocity in the direction of the drive,
$\langle V\rangle = N_v^{-1}\sum^{N_{v}}_{i} {\bf v}_{i}\cdot {\bf {\hat x}}$,
where ${\bf v}_i$ is the velocity of vortex $i$, and 
define the critical depinning force
$f_c$ as the value of $F^L$ at which
$\langle V\rangle=0.01$.
We have found that for slower sweep rates of the driving force,
there is no change in the measured depinning force.  
Finally, we note that our model should be valid for
stiff  vortices in three-dimensional superconductors
interacting with arrays of columnar defects. For a strictly 
two-dimensional
superconductor with periodic pinning arrays the vortex-vortex interaction
is modified from a Bessel function to $ln(r)$; however, our previous
simulations have indicated that either interaction produces
very similar commensurability effects 
\cite{Reichhardt,Jensen,Zimanyi,Olson}.     

\begin{figure}
\includegraphics[width=3.5in]{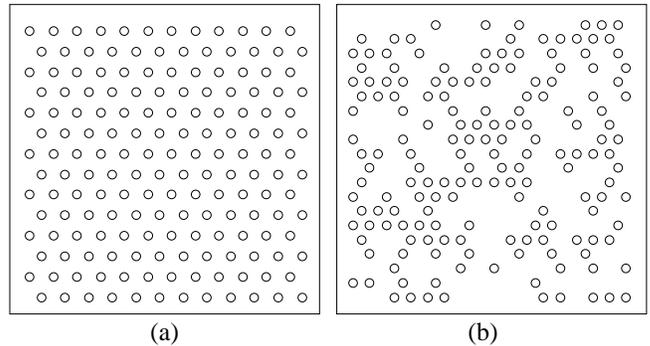}
\caption{
Circles: Pinning site locations for two 
$18\lambda \times 18\lambda$ samples with the same number 
of pinning sites 
$N_p=168$. 
(a) A triangular pinning array with no dilution, $P_d=0$, with
$B_\phi=0.52/\lambda^2$ and $B_\phi^*=0.52/\lambda^2$. (b) 
A triangular pinning array that has been diluted at $P_d=0.5$, with
$B_\phi=0.52/\lambda^2$ and $B_\phi^*=1.04/\lambda^2$. 
}
\end{figure}

\begin{figure}
\includegraphics[width=3.5in]{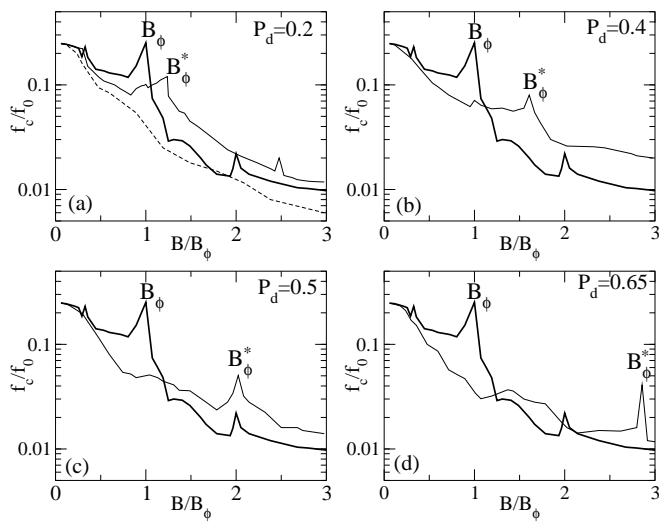}
\caption{The critical depinning force $f_{c}/f_{0}$ versus $B/B_{\phi}$
for systems with $B_{\phi} = 0.52/\lambda^2$.
The actual matching field $B_\phi$ and the matching field for an equivalent
undiluted array $B_\phi^*$ are labeled.
(a) $f_{c}/f_{0}$ for an undiluted pinning array with $P_d=0$ (heavy line), 
a pinning array with $P_{d} = 0.2$ (light line), 
and a random pinning arrangement (dashed line). 
(b) 
$P_d=0$ (heavy line) and 
$P_{d} = 0.4$ (light line). 
(c) 
$P_d=0$ (heavy line) and 
$P_{d}  = 0.5$ (light line). 
(d) 
$P_d=0$ (heavy line) and 
$P_{d} = 0.65$ (light line). 
}
\end{figure}

\section{Increasing Dilution}

\subsection{Fixed Pinning Density} 

In Fig.~1(a) we illustrate the positions of the pinning sites
for an $18\lambda \times 18\lambda$ system 
with no dilution, $P_d=0$.
The pinning sites are placed in a triangular array
at a pinning density of $B_{\phi} =  0.52/\lambda^2$,
giving $B_\phi=B_\phi^*=0.52/\lambda^2$. 
Figure~1(b) shows a system with $B_\phi=0.52/\lambda^2$ and
dilution $P_d=0.5$, so that $B_\phi^*=1.04/\lambda^2$.
Here, half of the pinning sites that would have formed a denser triangular
array have been removed randomly to give the same matching field $B_\phi$ as
in Fig.~1(a).

To illustrate the effect of the dilution,
in Fig.~2(a) we plot the critical depinning force
$f_{c}/f_{0}$ vs $B/B_{\phi}$ for three
systems with 
$B_\phi=0.52/\lambda^2$.
Two of the samples contain triangular
pinning arrays at different dilutions, $P_d=0$ [as in Fig.~1(a)] and
$P_d=0.2$.  The third sample contains randomly placed pinning.
For the undiluted triangular pinning array with $P_d=0$,
peaks in the depinning force occur at integer values of
$B/B_{\phi}$.  A submatching peak at
$B/B_{\phi} = 1/3$ also appears in agreement with 
earlier studies \cite{Jensen}. 
The random pinning arrangement shows no commensurability peaks and has a 
lower critical depinning force 
than the periodic pinning array for all but the very
lowest values of $B/B_\phi$.
The diluted periodic pinning array has a pronounced peak 
in $f_{c}/f_0$ 
at $B/B_{\phi} = 1.25$. This peak corresponds to the matching field
$B_{\phi}^{*}$ of the equivalent undiluted pinning array.
A second peak in $f_c/f_0$ occurs at $B=2B^{*}_{\phi}$.  There is a still a
small peak in $f_c/f_0$ at $B/B_{\phi}=1$ for the diluted array; however, 
we do not find a peak at $B/B_{\phi}=2$. 
The diluted pinning array has a higher $f_c/f_0$ than both the random pinning
arrangement and the
undiluted periodic pinning array 
for $B/B_\phi \gtrsim 1$ over the range of $B/B_\phi$ investigated here. 
In most of this range,
$f_{c}/f_0$ for the diluted periodic array is twice that of the 
undiluted periodic array. 

The peak in $f_c/f_0$ at $B=B^{*}_\phi$ persists as the dilution increases.
Figure 2(b) shows that a diluted pinning array with $P_d=0.4$ has a small
peak at $B/B_{\phi}=1$
and a more prominent peak at $B=B^{*}_{\phi}$. 
The critical depinning force is again higher in the diluted array 
than in the undiluted array for $B/B_{\phi} \gtrsim 1.1$. 
For a diluted pinning array with $P_d=0.5$, Fig.~2(c) indicates that the
commensurability peak in $f_c/f_0$ has shifted to $B=B^{*}_{\phi}=2B_\phi$.
Here, the pinning has become so dilute that 
there is no longer a noticeable enhancement of $f_c/f_0$
at $B/B_{\phi}=1$. 
When $P_d=0.65$, as in Fig.~2(d),
the 
peak in $f_c/f_0$ shifts up to $B=B^{*}_\phi=2.86 B_{\phi}$.
As $B^{*}_{\phi}$ increases with increasing dilution,
the critical depinning force at lower fields 
$B<B^{*}_\phi$ decreases. 
These results suggest that diluted periodic arrays may be useful
for increasing the overall pinning force at higher fields, and that 
a peak in the critical current at a specific field $B_p$ can be achieved 
using a {\it significantly smaller number of pins} than would be required to
create an undiluted periodic pinning array with $B_\phi=B_p$.

\begin{figure}
\includegraphics[width=3.5in]{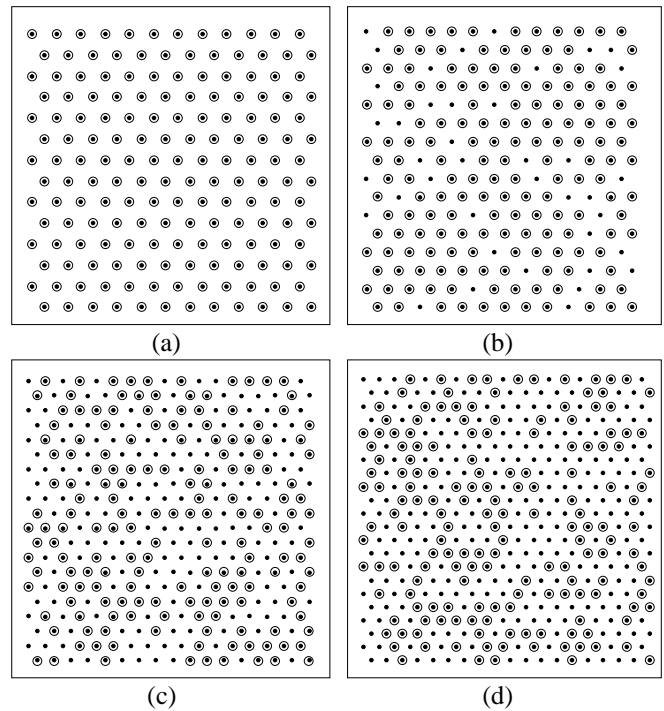}
\caption{Positions of pinning sites (open circles) and vortices (black dots)
for samples with $B_\phi=0.52/\lambda^2$ at $B/B^{*}_\phi=1$.
(a) $P_d=0$, $B^*_\phi=B_\phi$.
(b) $P_{d} = 0.2$, $B^{*}_{\phi} = 1.25B_\phi$.
(c) $P_{d} = 0.5$, $B^{*}_{\phi} = 2B_\phi$. 
(d) $P_{d} = 0.65$, $B^{*}_{\phi} = 2.86B_\phi$.    
}
\end{figure}

To understand the origin of the peak in $f_{c}/f_{0}$ 
at $B^{*}_{\phi}$, we analyze the vortex configurations 
at $B^*_\phi$ for four different values of $P_d$ in Fig.~3.  
For an undiluted triangular pinning array at $P_d=0$, 
as shown in Fig.~3(a), $B^*_\phi=B_\phi$,
each pin captures exactly one vortex, and there are no interstitial
vortices trapped in the regions between pinning sites.
In Fig.~3(b), at $B=B^*_\phi=1.25B_\phi$ in a system with $P_d=0.2$,
all of the pinning sites are occupied and the excess interstitial vortices
sit at the locations of the missing pinning sites, creating  a triangular
vortex lattice.
As $P_d$ increases, the increasing number of interstitial vortices present
at $B=B^*_\phi$ continue to occupy the locations of the missing pinning sites.
This is illustrated for $P_d=0.5$ in Fig.~3(c) and $P_d=0.65$ in Fig.~3(d).
The high symmetry of the triangular vortex lattice configuration at the
matching field $B=B^*_\phi$ causes the vortex-vortex interactions to cancel,
and the depinning force is determined primarily by the maximum force of the
pinning sites, $f_p$.
Defects in the vortex lattice appear just above and below 
$B=B^*_\phi$, creating asymmetrical vortex-vortex interactions.
A vortex associated with a defect experiences an extra force 
contributing to depinning on
the order of $F^{vv}(a/\lambda)$,
where $a$ is the vortex lattice constant, and $f_c$ is reduced.
A similar disordering process occurs just above and below each of the higher
matching fields 
\cite{Reichhardt},
although in this case the depinning force
is not merely determined by the pinning force since a portion of the vortices
are pinned indirectly in interstitial positions and experience a weaker
effective pinning force.

Previous work on wire network arrays \cite{Behrooz}
showed that even if the network is partially disordered, 
spatial correlations remain present as indicated by the appearance of
peaks in $k$-space and can produce matching effects.
In the diluted pinning arrays we consider here, even though a portion of the
sites are removed there are still peaks in the reciprocal space 
corresponding to the original undiluted array, so matching effects occur 
when the vortex lattice $k$ spacing is the same as these
pinning array $k$ space peaks. 
We note that if an equivalent number of randomly placed pinning sites 
is used there is no matching 
effect since there is a ring rather than peaks in $k$ space.

In the case of the diluted periodic pinning array, 
the vortex configuration contains numerous topological defects 
at $B=B_{\phi}$ for higher values of $P_d$ when the pinning is strong
enough that most of the vortices are trapped at the pinning sites.  
In contrast, at $B=B^{*}_{\phi}$ the vortex lattice is free of defects,
as illustrated in Fig.~3.
For fields just above or below $B=B^{*}_{\phi}$, vacancies and interstitials
appear in the vortex lattice and lower the depinning force.
At higher fields $B=nB^{*}_{\phi}$, where $n>1$ is an integer, 
the vortex lattice is again ordered
and a peak in the critical depinning force occurs, as shown
in Fig.~2(a) at $B=2B^{*}_{\phi}$.   

The enhancement of $f_{c}/f_{0}$ in the diluted
periodic pinning arrays for $B/B_\phi>1$ at the nonmatching fields 
occurs due to a different mechanism. 
We first note that at $B/B_\phi<1$, $f_c/f_0$ is lower in the diluted pinning
arrays
than in the undiluted pinning arrays
since it is possible for some of the vortices to sit in interstitial sites
instead of in pinning sites, even though not all of the pins have been
filled.
Interstitial vortices experience an effective pinning force due to the
interactions with the surrounding vortices.
Since the interstitial vortices are more weakly pinned
than vortices in pinning sites, 
the
critical depinning force is lower than it would be for an undiluted array 
at $B/B_\phi<1$, where every vortex sits in a pinning site.
For $B/B_\phi>1$, the initial motion of the vortices at depinning in an
undiluted array occurs via
channeling of the interstitial vortices between the pinning sites
along an effective modulated one-dimensional potential created by
the vortices located
at the pinning sites 
\cite{Zimanyi}.
In a diluted periodic pinning array at $B_\phi < B < B^*_\phi$, 
any channels of
interstitial vortices are interrupted by one or more pinning sites which
serve to increase the depinning threshold of the entire channel
by ``jamming'' free motion along the channel.  Here,
the one-dimensional potentials that allow for relatively free interstitial
vortex motion do not form until $B>B_\phi^*$ when the vortices begin to
occupy positions that would be interstitial sites in the equivalent
undiluted pinning array.
This picture is confirmed by our analysis of the vortex trajectories. 
The jamming
effect eventually breaks down at high dilutions $P_d$
when large regions devoid of pinning sites 
span the entire system and provide
macroscopic channels for free vortex flow. 
In Fig.~2(d) at $P_{d} = 0.65$, $f_{c}/f_0$ at
$B/B_{\phi}>1$ is lower than the $f_c/f_0$ values obtained at lower $P_{d}$
and similar fields. 
For higher values of $P_{d}$ this
effect becomes more pronounced until
$f_{c}/f_0$ eventually drops below the critical depinning current for the 
undiluted array.

\begin{figure}
\includegraphics[width=3.5in]{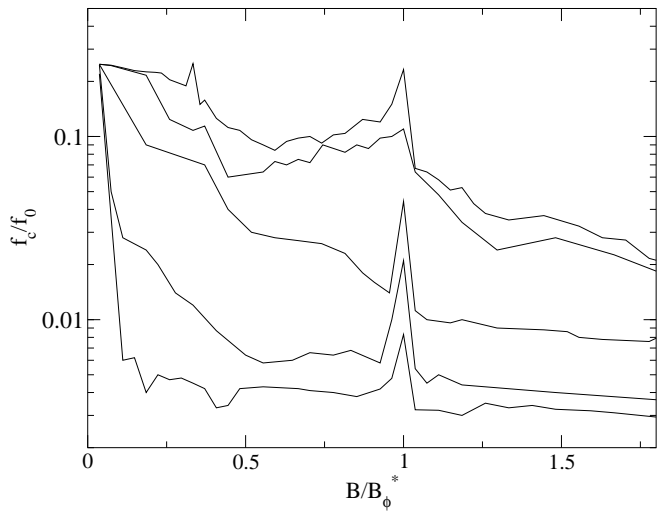}
\caption{$f_{c}/f_{0}$ vs $B/B^*_{\phi}$, where 
$B^*_{\phi}=0.83/\lambda^2$. 
Top curve: an undiluted periodic pinning array with $B_\phi=0.83/\lambda^2$.
Remaining curves, from top to bottom: Successive dilutions of the
same array to $P_d=0.25$, 0.5, 0.75, and 0.9.
}
\end{figure}

\begin{figure}
\includegraphics[width=3.5in]{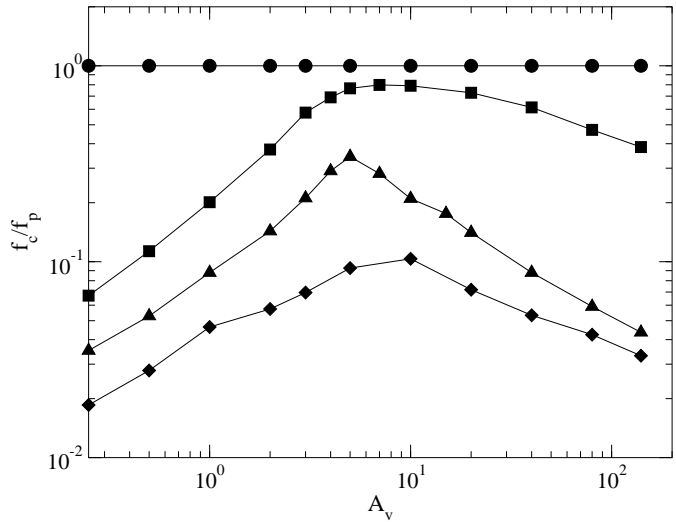}
\caption{$f_{c}/f_{p}$ vs $A_{v}$, the vortex-vortex interaction prefactor,
for four samples with $B=0.52/\lambda^2$.
Circles: undiluted triangular array with $B/B_{\phi}=1$
and $B_{\phi} = 0.52/\lambda^2$. 
Squares: diluted pinning array with $P_{d} = 0.5$ at $B/B^{*}_{\phi}=1$
where $B_{\phi}^* = 0.52/\lambda^2$. 
Triangles: undiluted array with 
$B/B_\phi=2.0$ and $B_{\phi} = 0.26/\lambda^2$. 
Diamonds: random array with
$B/B_\phi = 2.0$ and $B_{\phi} = 0.26/\lambda^2$.      
}
\end{figure}
      
\subsection{Decreasing Pinning Density} 
 
In Fig.~4 we show the effect of gradually diluting a periodic pinning array
with $B^*_\phi=0.83/\lambda^2$.
The main peak in the critical depinning current always falls at $B=B^*_\phi$,
but as the dilution increases from $P_d=0$ to $P_d=0.9$, the value
of $f_c/f_0$ decreases significantly.
The pronounced peak in $f_c/f_0$ persists even up to 
90\% dilution.
There is a small peak at $B=B_{\phi}$ for $P_d=0.25$,
but for higher
dilutions we find no noticeable anomalies in $f_c/f_0$ at $B=B_{\phi}$. 
This result 
indicates that the commensurability effect 
at $B=B^{*}_{\phi}$ is remarkably robust. 

\section{Effect of Vortex Lattice Stiffness}

We next consider the effect of the vortex lattice stiffness.
This parameter can be changed artificially
by adjusting the value of the factor $A_{v}$ in the vortex-vortex 
interaction term of Eq.~(2).
For dense random pinning where there are more pinning sites than 
vortices, $N_p>N_v$,    
decreasing the strength of the vortex-vortex interactions 
by lowering $A_{v}$ makes the vortex lattice softer and the
depinning force increases.  In the limit
of $A_{v} = 0$ the vortices respond as independent particles
and the depinning force will equal $f_{p}$ \cite{NE,Zimanyi2}.  
In the case where there
are more vortices than pinning sites 
$N_v>N_p$, the situation can be more complicated
since there are two types of vortices.  One species is located
at the pinning sites and the other species is in the interstitial sites. The
interstitial vortices are not directly pinned by the pinning sites but 
are restrained through the vortex-vortex interaction with pinned vortices. 
In this case it could be expected that
increasing the vortex-vortex interactions by raising $A_v$ would 
increase the depinning force.
On the other hand, if the vortex-vortex interaction force is strong enough, 
the interstitial vortices can more readily push the pinned vortices
out of the individual pinning sites.
Studies of very diluted random pinning arrays found that 
as the vortex-vortex interaction is increased, the depinning
force initially increases linearly; however, for high enough vortex-vortex
interactions the depinning curve begins to decrease again \cite{Fertig2}. 
The peak in the
depinning force coincides with a crossover from plastic depinning at 
low vortex-vortex interaction strengths to 
elastic depinning at high vortex-vortex
interaction strengths. A similar behavior was observed for  
vortices in periodic pinning where there were more vortices 
than pinning sites \cite{Zimanyi2}. 

For the results presented in Fig.~2 and Fig.~4, the vortex-vortex interaction
prefactor was set to $A_v=1.0$ 
and the depinning was plastic with the interstitial vortices
depinning first. In Fig.~5 we plot $f_{c}/f_{p}$ vs $A_{v}$ at
fixed $B=0.52/\lambda^2$ for 
four different pinning geometries. 
In an undiluted pinning array with $B/B_{\phi}=1$, 
the vortex lattice matches exactly with the
pinning array and the vortex-vortex interactions completely cancel.
As a result, the system responds like a single particle and 
$f_{c} = f_{p}$ for all $A_v$, as shown by the circles in Fig.~5.
When the same pinning array is diluted to $P_d=0.5$ 
and held at $B/B^*_\phi=1$, the squares in Fig.~5 illustrate that
for $ 0.25 \leq A_{v} < 5.0$ 
the depinning is plastic and 
$f_{c}/f_{p}$ linearly increases with $A_{v}$. 
This behavior is agreement with earlier studies \cite{Fertig2,Zimanyi2}.
For $A_{v} > 5.0$ the depinning is elastic and the entire lattice moves 
as a unit. In this elastic regime
$f_{c}/f_{p}$ first levels off and then decreases with increasing $A_v$ 
as the force exerted on the pinned vortices by the interstitial vortices
increases.
This result indicates that 
$f_c/f_p$ is maximized at the transition between plastic and elastic 
depinning. 
The filled triangles in Fig.~5 show the behavior of 
an undiluted array at $B/B_\phi=2$
where the vortex
lattice has a honeycomb structure rather than triangular symmetry 
\cite{Reichhardt}. 
Here the $f_{c}/f_{p}$ versus $A_v$ curve has a very similar trend to the
$P_d=0.5$ case, with a peak in $f_c/f_p$ at the crossover between plastic and
elastic depinning followed by a decrease in the depinning force for
$A_v>5.0$;
however, 
$f_c/f_p$ drops off more quickly for the
undiluted pinning array at $B/B_\phi=2$
than for the diluted pinning array with $P_d=0.5$.
The vortex lattice is always triangular for the $P_d=0.5$ diluted pinning array
at $B/B^{*}_{\phi}=1$, so as $A_{v}$ increases there is no change in the 
vortex lattice structure and every pinning site remains occupied. 
For the undiluted pinning array at $B/B_\phi=2$ the vortex lattice 
experiences an increasing amount of distortion for $A_{v} > 5.0$  as
the structure changes from a honeycomb arrangement in which all of the
pinning sites are occupied to a triangular arrangement in which some pins
are empty.
This implies that for stiff vortex lattices,
a relatively larger depinning force can be obtained from a diluted pinning
array than from an undiluted pinning array with an equal number of pinning
sites.  
The triangles in Fig.~5 show the behavior of a random pinning array 
at a pinning density of $B_\phi=0.26/\lambda^2$
with $B/B_\phi = 2$. 
We find the same trend of a peak in $f_c/f_p$ as a function of $A_v$;
however, $f_c/f_p$ for the random pinning array is always less than 
that of the periodic or diluted pinning arrays.
We note that a peak in $f_c/f_p$ still appears for vortex densities 
that are well away from commensuration due to the competition between the
pinning energy and elastic vortex lattice energy.

One aspect we did not explore in this work 
is the effect of melting and thermal fluctuations on the diluted pinning
arrays. 
Melting would be relevant in strongly layered superconductors with periodic
arrays of columnar defects. If a diluted periodic pinning array 
could be fabricated in a layered superconductor, it is likely that 
a rich variety of thermally induced effects would appear.
One system that should have similar behavior is a sample with a low
density of randomly placed columnar defects in 
which the number of vortices is much
greater than the number of pinning sites \cite{Menghini}. 
Theoretical 
work suggests that such a system can 
undergo a Bragg-Bose glass transition \cite{Valls,Dasgupta}.
In the case where the pinning is periodic, it would be interesting to 
determine whether the melting transition shifts upward or downward 
at $B/B^{*}_\phi=1$ compared to away from $B/B^{*}_\phi=1$. 
Additionally, at intermediate dilution it may be possible
to observe two-stage melting or a nanoliquid. 
Such a nanoliquid has been observed in 
experiments with patterned regions of columnar defects 
\cite{Banerjee}.       

\section{Summary} 

In summary, we have demonstrated that commensuration 
effects can occur  in superconductors with diluted periodic pinning arrays. 
Pronounced peaks in the critical depinning current 
occur at magnetic fields corresponding to the matching field for the 
undiluted array, $B^*_\phi$, which 
may be significantly higher than the matching
field $B_\phi$ for the pinning sites that are actually present in the sample.
The commensurability effect is associated with 
the formation of ordered vortex
lattice arrangements containing no topological defects. 
This effect arises since the diluted pinning array has
correlations in reciprocal space at the same values of $k$ as the
undiluted pinning array, enabling a matching effect to occur for
the diluted array when the vortex density matches the density of 
the undiluted array.
The effect is remarkably robust, and
peaks in the critical depinning force persist in pinning arrays
that have been diluted up to 90\%.  
For small dilutions, a peak can also appear at the true matching field
$B_\phi$.
In samples with equal numbers of pinning sites, diluted periodic arrays
have smaller critical depinning currents than undiluted arrays 
at fields below $B_\phi$, but produce a significant enhancement of the
critical current at fields above $B_\phi$ compared to both undiluted arrays
and random pinning arrangements.
This enhancement for $B < B_{\phi}$ is due to 
the suppression of easy channels of one-dimensional motion that 
occur in the undiluted arrays.
We have also examined the effect of the vortex lattice stiffness
on depinning. 
For $B>B_\phi$, the depinning is plastic in soft lattices 
and the interstitial vortices begin to move before the pinned vortices,
while for stiff lattices the depinning is elastic and
the entire lattice depins as a unit.
In the plastic depinning regime
which appears at small lattice stiffness,
the depinning force increases with increasing lattice stiffness
since the interstitial vortices become
more strongly caged by the vortices at the pinning sites. 
In the elastic depinning regime the depinning force decreases with 
increasing lattice stiffness.
A peak in the depinning force occurs at the crossover between
plastic and elastic depinning. 
For stiffer lattices the diluted pinning arrays show more pronounced
matching effects. 
Our results suggest that diluted periodic pinning arrays may be
useful for enhancing the critical current at high fields in systems where only
a limited number of pinning sites can be introduced.

This work was carried out under the auspices of the 
NNSA of the U.S. Dept.~of Energy at 
LANL under Contract No.~DE-AC52-06NA25396.

\end{document}